\documentclass[11pt]{llncs}
\usepackage{amsfonts}
\usepackage{times}
\usepackage{color}
\usepackage{graphicx,epsfig}
\usepackage{wrapfig}
\newcommand{\RR}{\mathbb{R}}

\newcommand{\argmax}{\arg\max}

\newcommand{\commentout}[1]{}

\title{Computing an Aggregate Edge-Weight  Function for Clustering Graphs with Multiple Edge Types\thanks{This work is supported by the Laboratory Directed Research and Development program of Sandia National Laboratories. }}
\author{ Matthew Rocklin\inst{1} \and Ali Pinar\inst{2}}
\institute{\email{mrocklin@cs.uchicago.edu} Department of Computer Science, University of Chicago \and \email{apinar@sandia.gov} \footnote{ This author is also supported by  DOE Applied Mathematics Research Program} Sandia National Laboratories\footnote{Sandia National Laboratories is a multi-program laboratory operated by Sandia Corporation, a wholly owned subsidiary of Lockheed Martin Corporation, for the U.S. Department of EnergyÕs National Nuclear Security Administration under contract DE-AC04-94AL85000.}}

\begin{document}
\maketitle
\begin{abstract}
We investigate  the community detection problem on graphs in the existence of multiple edge types.  Our main motivation is that similarity between objects can be defined by many different metrics and  aggregation of  these metrics into a single one poses several important challenges, such as recovering this aggregation function from  ground-truth, investigating the space of different clusterings, etc.  In this paper, we address  how to  find an aggregation  function to generate
a composite metric that best resonates with the ground-truth.  We describe two approaches: solving an inverse problem where we try to find parameters  that generate a graph whose clustering gives the ground-truth clustering, and choosing parameters to maximize the quality of the ground-truth clustering.   
We present experimental results on real and synthetic   benchmarks. 
\end{abstract}

\vspace{-5ex}
\section{Introduction}
\label{sec:introduction}
\vspace{-2ex}

 A community or a cluster in a network is assumed to be a subset of vertices that are tightly coupled among themselves and loosely coupled with the rest of the network.    Finding these communities is one of the fundamental problems of networks analysis and  has  been the subject of numerous research efforts.   
 Most of these efforts begin with the premise that a simple graph is already constructed. That is  the  relation between two objects (hence existence of a single edge between two nodes) is already quantified with a binary variable or a single number that represents the strength of the connection.  This paper studies the community detection problem on networks with multiple edges types or multiple similarity metrics, as opposed to traditional networks with a single edge type.   

In many real-world problems,  similarities between objects can be defined by  many different relationships. For instance,  similarity between two  scientific articles can be defined based on authors, citations to, citations from, keywords, titles, where they are published,  text similarity and many more.  Relationships between people can be based on  the nature of communication (e.g., business, family, friendships) or the means of communication (e.g., emails, phone, in person).  Electronic files can be  grouped  by their type (Latex, C,  html),  names, the time they are created,  or the pattern they are accessed. In these examples, there are actually multiple graphs that define  relationships  between the subjects.  Reducing all this information by constructing a single composite graph is convenient as it enables application of many strong results from the literature. However, the information being lost  during this aggregation may be crucial.   

The community detection problem on networks with multiple edge types bears many interesting problems.  If the ground-truth  clustering is known, can we recover an aggregation scheme that best resonates with the ground-truth data? 
Is there a meta-clustering structure, (i.e., are the clusterings clustered) and how do we find it?  How do we find significantly different clusterings for the same data? 
%
%
These problems add another level of complexity to the already difficult problem of community detection in networks.  As in the single edge type case,  the challenges lie not only in  algorithms, but also  in formulations of these problems.  
Our ongoing work addresses all these problems. In this paper however, we will focus on  recovering an aggregation scheme  for  ground-truth clustering. 
Our techniques rely on using nonlinear optimization and   methods for classical community detection (i.e., community detection with single edge types). We present results with real data sets as well as synthetic data sets. 

\vspace{-2ex}
\section{Background}
\label{sec:background}
\vspace{-2ex}
Traditionally, a graph $G$ is defined as a tuple $(V,E)$, where $V$ is a set of vertices and $E$ is a set of edges. A weight $w_i\in \RR$ may be associated with the edges that corresponds to the strength of the connection between the two end vertices. In this work, we work with multiple edge types that correspond to different measures of similarity.   Subsequently, we  replace the weight of an edge $w_i\in \RR$  with a weight vector $\langle w_i^1,w_i^2,\ldots ,w_i^K\rangle \in \RR^K$, where $K$is the number of  different edge types.  A composite similarity can be defined by a function $\RR^K\rightarrow \RR$  to reduce the  weight vector to a single number.  In this paper, we will restrict ourselves to linear functions such that 
 the composite edge weight $w_i(\alpha)$ is defined as $\sum_{j=1}^K\alpha_jw_i^j$.



\subsection {Clustering in graphs}

Intuitively, the goal of clustering is to break down the graph into smaller groups such that   vertices in each group  are tightly coupled among themselves, and loosely coupled with the remainder of the network.   Both the translation of this intuition into a well-defined mathematical formula and design of associated algorithms pose big challenges.  
Despite the  high quality and the high volume of the literature, the area continues to draw a lot of interest  both due to the growing importance of the problem and the challenges  posed by the sizes of the subject graphs and the mathematical variety as we get into the details of these problems.   

Our goal here is to extend the concept of clustering to graphs with multiple edge types without getting into the details of clustering algorithms  and formulations, since such a detailed study will be well beyond the scope of this paper. 
In this paper, we used {\it Graclus}, developed by Dhillon et al\cite{Dhillon2007a}, which uses  the top-down approach that recursively  splits the graph into smaller pieces. 



\subsection {Comparing two clusterings}
At the core of most of our discussions will be  similarity between two clusterings.
Several metrics and methods  have been proposed for comparing clusterings, such as  {\it variation of information}~\cite{meila05}, {\it scaled coverage measure}~\cite{stichting00}, {\it classification error}~\cite{lange04,luo05,meila05}, and {\it Mirkin's metric}~\cite{mirkin96}.  Out of these, we have used the variation of information metric in our experiments.


Let $C_0=\langle C_0^1,C_0^2, \ldots ,C_0^K\rangle$ and $C_1=\langle C_1^1,C_1^2, \ldots ,C_1^K\rangle$ be two clusterings of the same node set.  
Let $n$ be the total number of nodes, and  $P(C, k)=\frac{|C^k|}{n}$ be the probability that a node is in cluster $C^k$ in a clustering $C$.  We also define $P(C_i,C_j,k,l)= \frac {|C_i^k\cap C_j^l|}{n}$. 
Then the {\it entropy of information} in  $C_i$  will be 
\[ H(C_i)=-\sum_{k=1}^K P(C_i,k)\log{P(C_i,k)}  \] 
the mutual information shared by $C_i$ and $C_j$  will be
\[ I(C_i,C_j)=\sum_{k=1}^K\sum_{l=1}^{K}P(C_i,C_j,k,l)\log {P(C_i,C_j,k,l)},
\]
%
and the variation of information is given by
\begin{equation} 
\label{eq:vi1}
d_{VI}(C_i,C_j)=H(C_i)+H(C_j)-2I(C_i,C_j).  
\end{equation}
Meila~\cite{meila05} explains the intuition behind this metric a follows.  $H(C_i)$ denotes the average uncertainty of the position of a node in  clustering $C_i$.  If, however, we are given $C_j$, $I(C_i,C_j)$ denotes average reduction in uncertainty of where a node is located in $C_i$. If we rewrite Equation (\ref{eq:vi1}) as 
 \[ 
d_{VI}(C_i,C_j)=\left(H(C_i)-  I(C_i,C_j)\right)\;\; +\;\; \left(H(C_j)-  I(C_i,C_j)\right),
\]
the first term will be  measurement of information lost if $C_j$ is the true clustering and we obtain $C_i$, and the second term will be vice versa. 
 
The variation of information metric can be computed in $O(n)$ time. 
 



\vspace*{-3ex}
\section{Recovering a graph given a ground truth clustering}
\label{sec:groundtruth}
\vspace{-1ex}


Suppose we have the ground-truth clustering information about a graph with multiple similarity metrics.
Can we recover an aggregation scheme that best resonates with the ground-truth data?  This aggregation scheme that reduces multiple similarity measurements into a single similarity measurement can be  a crucial enabler that reduces the problem of  finding communities with multiple similarity metrics,  to a well-known, fundamental problem in data analysis. Additionally if we can obtain this aggregation scheme from data sets for which the ground-truth is available, we may then apply the
same aggregation to other data instances in the same domain. 

Formally,  we work on the following problem. Given a graph $G=(V,E)$ with multiple  similarity measurements for each edge $\langle w_i^1,w_i^2,\ldots ,w_i^K\rangle \in \RR^K$, and a ground-truth clustering for this graph $C^*$. 
Our goal is to find a weighting vector $\alpha\in \RR^K$, such that the $C^*$ is an optimal clustering for the graph $G$, whose edges are weighted as $w_i=\sum_{j=1}^K\alpha_jw^j_i$.   Note that this is only a semi-formal definition, as we have not formally defined what we mean by an {\it optimal clustering}.   In addition  to the well-known difficulty of  defining what a good clustering means, matching to the ground-truth data has specific challenges, which  we discuss in the subsequent section. 
 




Below, we describe two approaches. The first  approach  is based on inverse problems, and we try to find weighting parameters for which  the clustering on the graph yields the ground-truth clustering.  The second approach computes weighting parameters that maximizes the quality of the ground-truth clustering. 

\vspace{-1ex}
\subsection{ Solving an inverse problem}

\vspace{-1ex}
Inverse problems arise in many scientific computing applications where  the goal is  to infer unobservable parameters from finite observations.  Solutions typically  involve iterations of taking guesses   and then solving the {\it  forward} problems to compute the quality of the guess.  
Our problem can be considered as an inverse problem, since we are trying to  compute an aggregation function, from  a given clustering.  The forward problem in this case will be  the clustering operation.  We can start with a random guess for the edge weights, cluster the new graph, and use the distance between two clusterings as a measure for the quality of the guess. We can further put this process within an optimization loop to find the parameters that yield  the closest clustering to the ground-truth.  

The disadvantage of this method is that it relies on the accuracy of the forward solution, i.e., the clustering algorithm.  If we are given the true solution to the problem, can we construct the same clustering?  This will not be easy for two reasons. 
First,  there is no perfect clustering algorithm, and secondly, even if we were able to solve the clustering problem optimally, we would not  have the exact objective function for clustering. 
Also, the need to solve many clustering problems will be time-consuming especially for large graphs.  

%
\vspace{-2ex}
\subsection{Maximizing the quality of ground-truth clustering}
\vspace{-1ex}
An  alternative approach is to find an aggregation function that maximizes the quality of the ground-truth clustering.
For this purpose, we have to take into account not only  the overall quality of the clustering, but also the  placement of individual vertices, as the individual vertices represent local optimality.  For instance, if the quality of the clustering will improve by moving a vertex to another cluster than its  ground-truth, then the current solution  cannot be ideal.  While it is fair to assume some vertices might have been misclassified in the ground-truth data, there should be a penalty for such vertices. Thus we have two objectives while computing $\alpha$: {\it (i)} justifying the location of each vertex  
{\it (ii)}  maximizing the overall quality of the clustering. 
%
\vspace{-2ex}
\subsubsection{Justifying locations of individual vertices}







For each vertex $v\in V$ we define the \textit{pull} to each cluster $C^k$  in $C=\langle C^1,C^2, \ldots  C^K  \rangle $ to be the cumulative weights of edges between $v$ and its neighbors in $C^k$, 
\begin{equation}
P_\alpha(v,C_k) = \sum_{w_i=(u,v)\in E; u\in C^k} w_i(\alpha)
\end{equation}
We further define the \textit{holding power}, $H_\alpha(v)$ for each vertex, to be the pull of the cluster to which the vertex belongs in $C^*$ minus the next largest pull among the remaining clusters. If this number is positive then $v$ is held more strongly to the proper cluster than to any other. 
We can then maximize the number of vertices with positive holding power by maximizing $ |\{v : H_\alpha(v)>0\}|$.
What is important for us here is the concept  of pull and hold, as the specific definitions may be changed without altering the core idea.



While this method is local and easy to compute, its discrete nature limits the tools that can be used to solve the associated optimization problem. Because gradient information is not available it hinders our ability to navigate in the search space.  In our experiments, 
we smoothed the step-like nature of the function $H(v)>0$ by replacing it with $\arctan(\beta H_\alpha(v))$. This functional form still encodes that we want holding power to be positive for each node but it allows the optimization routine to benefit from small improvements. It emphasises nodes which are close to the $H(v)=0$ crossing point (large gradients) over nodes which are well entrenched (low gradients near extremes).

This objective function sacrifices holding scores for nodes which are safely entrenched in their cluster (high holding power) or are lost causes (very low holding power) for those which are near the cross-over point. The extent to which it does this can be tuned by $\beta$, the steepness parameter of the arctangent.\begin{wrapfigure}{l}{28ex}
\vspace{-2.5ex}
\hspace*{-5ex}
\includegraphics[width=.5\textwidth]{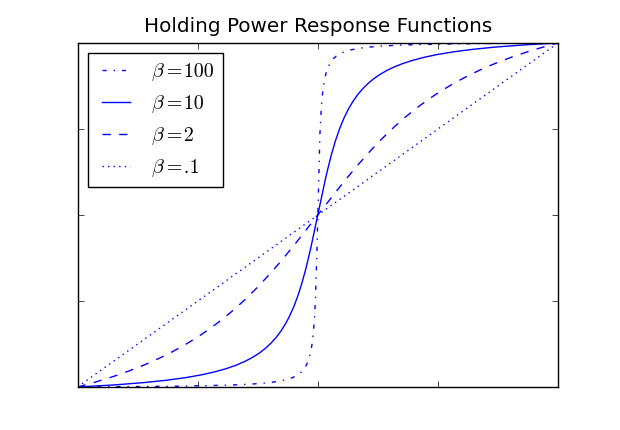}
\vspace{-4.5ex}
\caption{\label{fig:responsefunctions} Arctangent provides a smooth blend between step and linear functions. }
\vspace{-5ex}
\end{wrapfigure}
  For very steep parameters this function resembles classification (step function) while for very shallow parameters it resembles a simple linear sum as seen in
Fig.~\ref{fig:responsefunctions}. 
 We can solve the following optimization problem to maximize the number of vertices, whose positions in the ground-truth clustering are justified by the weighting vector $\alpha$.  
\begin{equation}
\argmax_{\alpha\in\RR^K} \sum_{v\in V} \arctan(\beta H_\alpha(v))
\end{equation}


\vspace{-2ex}
\subsubsection {Overall clustering quality}
\vspace{-1ex}

In addition to individual vertices  being justified, overall quality of the clustering should be  maximized. Any quality metric can potentially be used for this purpose however we find that some strictly linear functions have a trivial solution. Consider an objective function that measures the quality of a clustering as the sum of the inter-cluster edges. 
To minimize the cumulative weights of cut edges, or equivalently to maximize the cumulative weights of internal edges we solve
\begin{eqnarray*} 
 \min_\alpha &   \sum_{e_{j}\in Cut}\sum_{i=1}^K \alpha_i w^i_j \;\;\; |\alpha|=1\\
\end{eqnarray*}
where $Cut$ denotes the set of edges whose end points are in different clusters. 
Let $S^k$ denote the sum of the cut edges with respect to the $k$-th metric.
 That is $S^k=\sum_{e_{j}\in Cut} w_j^k$.  Then the objective function can be rewritten as $\min_\alpha \sum_1^K \alpha_kS^k$. Because this is linear it has a trivial solution that assigns 1 to the  weight of the maximum $S^k$, which means  only one similarity metric is taken into  account. 
While additional constraints may exclude this specific solution, a linear formulation of the quality will always yield only a trivial solution within the new feasible region.  

In our experiments we used the {\it modularity}  metric~\cite{modularity}.  The modularity metric  uses a random graph  generated with respect to the degree distribution as  the null hypothesis, setting the modularity score of a random clustering to 0.  Formally,  the modularity score for an unweighted graph  is 
\begin{equation}
\frac{1}{2m}\sum_{ij} \left[e_{ij} - \frac{d_id_j}{4m}\right]\delta_{ij},
\end{equation}
where $e_{ij}$ is a  binary variable that is 1, if and only if  vertices $v_i$ and $v_j$ are connected;  $d_i$ denotes the degree of vertex $i$; $m$ is the number of edges; and $\delta_{i,j}$  is a  binary variable that is 1, if and only  if  vertices $v_i$ and $v_j$ are on the same cluster.   In this formulation,  $\frac{d_id_j}{4m}$ corresponds to the number of edges between vertices $v_i$ and $v_j$ in a random graph with the given degree distribution, and its subtraction corresponds to the  the null hypothesis.

This formulation can be generalized for weighted graphs by redefining $e_{ij}$ as  the weight of this edge (0 if no such edge exists), $d_i$ as  the cumulative weight of edges incident to $v_i$; and $m$ as the cumulative weight of all edges in the graph~\cite{wmodularity}. 

 \vspace{-2ex}
\subsection{Solving the optimization problems }
\label{sec:opt}
\vspace{-2ex}
We have presented several nonlinear optimization problems for which the derivative information is not available. 
 To solve these problems we used HOPSPACK (Hybrid Optimization Parallel Search PACKage)~\cite{hopspack}, which is developed at Sandia National Laboratories to solve linear and nonlinear optimization problems  when the derivatives are not available.  
 


%
\vspace{-2ex}
\section{Experimental results}
\vspace{-2ex}
%



\subsection {Recovering edge weights}
\vspace{-1ex}
 The goal of this  set of experiments is  to see whether we can find aggregation functions that justify a given clustering. We have performed our experiments on 3 data sets.

\noindent{\bf  Synthetic data:}
Our generation method is based on Lancichinetti et al.'s work \cite{Lancichinetti2008} that proposes a method to generate graphs as benchmarks for clustering algorithms.
We generated networks of sizes 500, 100, 2000, and 4000 nodes, 30 edges per node on average, mixing parameters $\mu_t=.7, \mu_w=.75$, and known communities. We then perturbed edge weights, $w_i$, with additive and multiplicative noise so that $w_i\leftarrow \nu(w_i+\sigma) : \sigma\in (-2w_{a}, 2w_{a}), \nu \in(0,1)$ uniformly, independently, and identically distributed, where $w_{a}$ is the average edge weight.

\begin{figure}[bth]
\vspace{-4ex}
\centering
\includegraphics[width=1.0\textwidth]{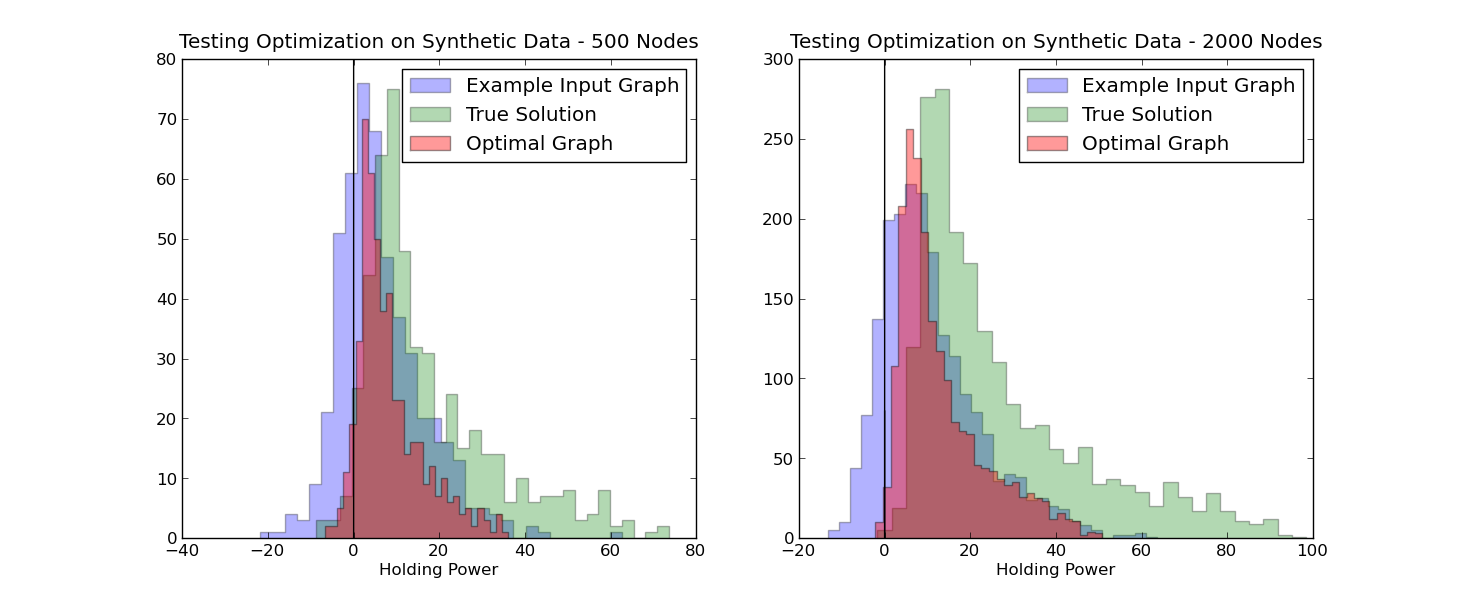}
\vspace{-3ex}
\caption{\label{fig:lfrhist} Three histograms of holding powers for Blue: an example perturbed (poor) edge type, Green: the original data (very good), Red: the optimal blend of ten of the perturbed edge types. }
\vspace{-4ex}
\end{figure}

After the noise, none of the original metrics  preserved the original clustering structure. We display this in Fig.~\ref{fig:lfrhist}, which  presents histograms for the holding power for vertices. The green bars correspond to vertices of the original graph, they all have positive holding power.  The blue bars correspond to holding powers after  noise is added. We only present  one edge type for  clarity of presentation. As can be seen a significant portion
(30\%) of the vertices have negative holding power, which means  they would rather be on another cluster.  The red bars show the holding powers after we compute an optimal linear aggregation. As seen in the figure, almost all vertices move to the positive side, justifying the ground-truth clustering.  A few vertices with negative holding power are expected, even after an optimal solution due to the noise. These  results show that a composite similarity  that resonates with
a given clustering can be computed  out of many metrics, none of which give a good solution by itself.  


In Table~\ref{tab:lfrtest},  we present these results on graphs with different number of vertices.  While the percentages change for different number of vertices, our main conclusion that a good clustering can be achieved  via a better  aggregation function remains valid. 

\begin{table}[tbh]
\vspace{-3ex}
\caption{ \label{tab:lfrtest}Fraction of nodes with positive holding power for  ground-truth, perturbed, and optimized networks}
\vspace{-2ex}
\begin{center}
  \begin{tabular}{| c | c | c | c | c | }
    \hline
    Number of nodes &Number of clusters&  Ground-truth & Optimized  & Perturbed (average) \\ \hline
    500 &  14  & .965 & .922 & .703 \\
    1000 & 27 & .999 & .977 & .799 \\
    2000 & 58 &  .999 & .996 & .846 \\
    4000 & 118& 1.00 & .997 & .860 \\
    \hline
  \end{tabular}
\end{center}
\vspace{-5ex}
\end{table}


\noindent
{\bf  File system data:} 
An owner of a  workstation classified  300 of his files as belonging to one of his three ongoing projects, which we took as the ground-truth clustering.  We used  filename similarity, time-of-modification/time-of-creation similarity, ancestry (distance in the directory-tree), and parenthood (edges between a directory node with file nodes in this directory) as the similarity metrics among these files. 

\begin{figure}[thb]
\vspace{-4ex}
\centering
\includegraphics[width=0.6\textwidth]{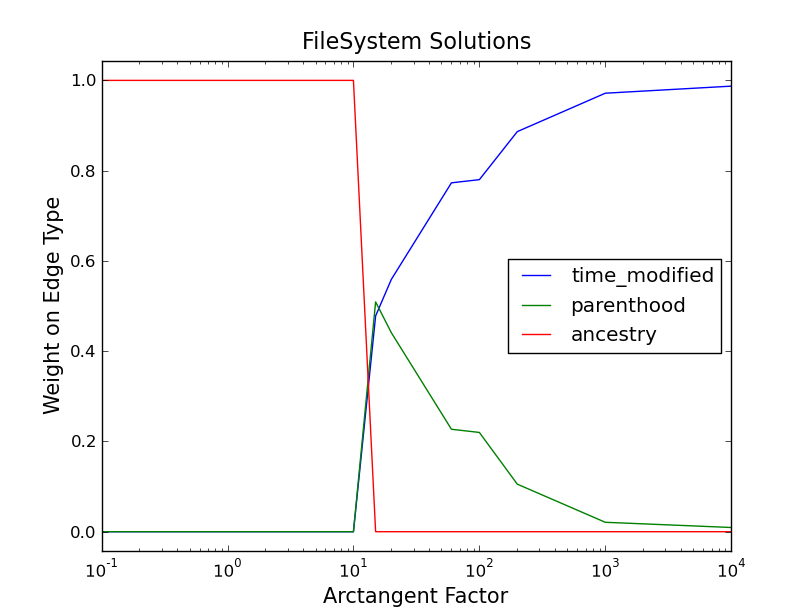}
\vspace{-2ex}
\caption{Optimal solutions  for the file system data for different arctangent parameters} 
\label{fig:filesystem}
\vspace{-4ex}
\end{figure}
Our  results showed that only three metrics (time-of-modification, ancestry, and parenthood) affected the clustering.  However, the  solutions were sensitive to the choice of the arctangent parameter. In Fig.~\ref{fig:filesystem}, each column corresponds to an optimal solution for the corresponding  arctangent parameter.  Recall that  higher values of  the arctangent parameter corresponds to sharper step functions. Hence, the right of the figure corresponds to  maximizing the total number of  vertices with positive holding power, while the left side corresponds to  maximizing the sum of holding powers. The difference between the two is that the solutions on the right side may have a lot of nodes with barely positive values, while those on the left  may have nodes further away from zero at the cost of  more nodes with negative holding power. 
This is expected in general, but drastic change in the optimal solutions  as we go from one extreme to another was surprising to us, and should be taken into account in further studies. 

\noindent{\bf Arxiv data:}
We  took  30,000 high-energy physics articles published on arXiv.org and considered abstract text similarity, title similarity, citation links, and shared authors as edge types for these articles. We used  the top-level domain  of the submitter's e-mail (.edu, .uk, .jp, etc...) as a proxy for the region where the work was done. We used these regions as the ground-truth clustering.

The best parameters  that  explained the  ground-truth clustering were  0.0 for  abstracts, 1.0 for authors, 0.059  for citations, and 0.0016 for titles.  This means the  shared authors edge type is almost entirely favored, with cross-citations coming a distant second. This is intuitive because a network of articles linked by common authors will be linked both by topic (we work with people in our field) but also by geography (we often work with people in nearby institutions) whereas edge types like abstract text similarity tend to encode only the topic of a paper, which is less geographically correlated.  Different groups can work on the same topic, and  it was good to see that citations  factored in, and such a clear dominance of the authors information was noteworthy.  As a future work, we plan to investigate  nonlinear aggregation functions on this graph. 

%


%
\vspace{-2ex}
\subsection{Clustering quality vs.  holding vertices } 
\begin{figure}[t]
\vspace{-4ex}
\centering
\includegraphics[width=.55\textwidth]{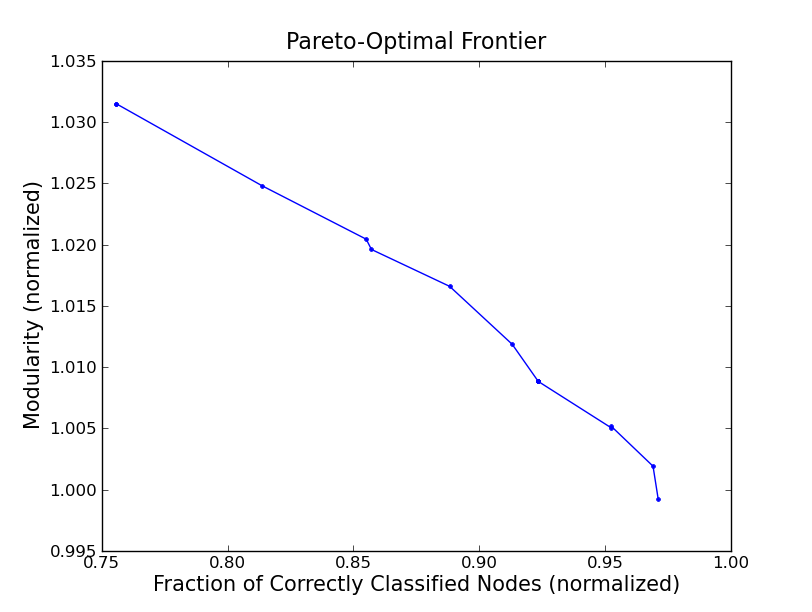}
\vspace{-2ex}
\caption{Pareto frontier for two objectives: normalized modularity  and percentage of nodes  with positive holding power}
\label{fig:pareto}
\vspace{-4ex}
\end{figure}

\vspace{-2ex}
We have stated two goals  while computing an aggregation function: justifying the position of each vertex and the overall quality of clustering.  
In Fig.~\ref{fig:pareto}, we present the Pareto frontier for the two objectives.  The vertical axis  represent the quality of the clustering with respect to the modularity metric~\cite{modularity}, while the horizontal axis represents the percentage of nodes with positive holding power.  The  modularity numbers are normalized with respect to the modularity of the ground-truth clustering, and  normalized  numbers can be above 1, since  the ground-truth clustering does not specifically aim at maximizing modularity. 

As expected, Fig.~\ref{fig:pareto}  shows a trade-off between two objectives.  However, the scale difference between the two axis should be noted.  The full range in modularity change is limited to only 3\% for modularity, while the range is more than 20\% for fraction of vertices with positive holding power. More importantly,  by only looking at the holding powers we can preserve the original modularity quality.   The reason for this is that we have relatively small clusters, and almost all vertices
 have a connection with a cluster besides their own. If we had clusters where  many vertices had all their  connection within their clusters (e.g., much larger clusters), then this would not have been the case, and having a separate quality of clustering metric would have made sense.  However, we know that most complex networks have small communities no matter how big the graphs are~\cite{leskovec:conductance}.  Therefore, we expect that looking only at the  holding powers of vertices will be sufficient to recover aggregation functions.  

 \vspace{-2ex}
 \subsection{Inverse problems  vs.  maximizing clustering quality} 
\vspace{-1ex}
\begin{figure}[t]
\vspace{-4ex}
\centering
\includegraphics[width=.6\textwidth]{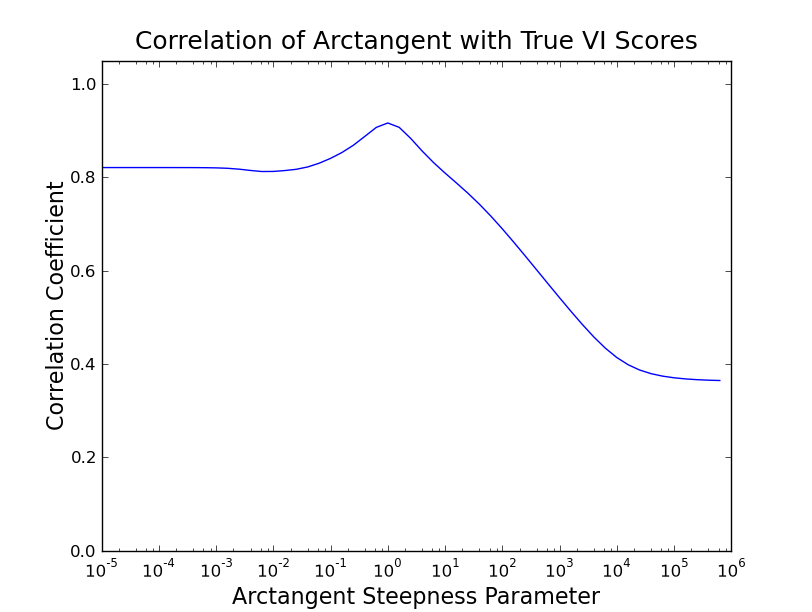}
\vspace{-2ex}
\caption{The correlation of the arctan-smoothed objective function with variation of information distance using clusterings generated by Graclus  as we vary the steepness parameter} 
\label{fig:correlation}
\vspace{-4ex}
\end{figure}

We used the file system data set to investigate the relationship between the two proposed approaches, and present results in Fig.~\ref{fig:correlation}.  For this figure we compute the objective function for the ground-truth clustering for various aggregation weights and use the same weights to compute clusterings  with Graclus. From these clusterings we compute the variation of information (VI) distance to the ground-truth.
Fig.~\ref{fig:correlation} presents the correlation between the  measures:  VI distance for Graclus clusterings for the first approach, and the objective function values for the  second approach.  
This tries to  answer whether solutions with higher  objective function values yield clusterings closer to the ground-truth using Graclus.   In this figure,  a horizontal line fixed at 1 would have shown complete agreement.  Our results show a strong correlation when moderate values for $\beta$ are taken (arctan function is neither too step-like nor too linear).  These results are not sufficient  to be conclusive as we need more experiments and other clustering tools.  However, this experiment produced promising results and  shows how such a study may be performed.   

%

%
\vspace{-2ex}
\subsection{ Runtime scalability}
\begin{wrapfigure}{l}{36ex}
\vspace{-4ex}
\centering
\hspace*{-4ex}
\includegraphics[width=.5\textwidth]{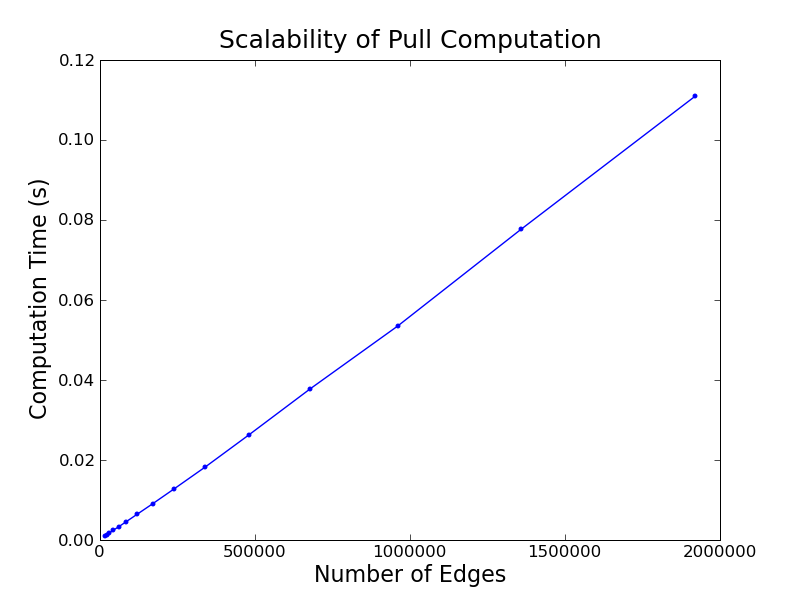}
\vspace{-2ex}
\caption{Scalability of the proposed method }
\label{fig:runtime}
\vspace{-4ex}
\end{wrapfigure}

\vspace{-1ex}
In our final set of experiments we show the scalability of the proposed method. First, we want to note that  the number of unknowns  for the optimization problem is only a function of the aggregation function and is independent of the  graph size.  The required number of operations for one function evaluation on the other hand  depends linearly on the size of the graph, as illustrated in Figure~\ref{fig:runtime}.  In this experiment, we used  synthetic graphs with 30 as the average degree, and the presented numbers correspond to averages on 10 different runs.   As expected, the runtimes scale linearly with the number of edges.   
     
The runtime of the optimization algorithm depends  on the number of function evaluations. Since the algorithm we used is nondeterministic, the number of function evaluations, hence runtimes vary  even for different runs on the same problem, and thus are less informative. We are not presenting these results in detail due to space constraints.  However, we want to reemphasize that the size of the optimization problem  does not grow with  the graph size, and we don't expect the number of functions evaluations to cause any scalability problems.   

 We also observed that the number of function evaluations increase linearly with the number of similarity metrics.  These results are also omitted due to space constraints. 

\section{Conclusion and Future Work}
\label{sec:conclusion}
We have discussed the problem of  graph clustering with  multiple edge types, and studied computing an aggregation function to compute composite edge weights that best resonate with given ground-truth clustering. 
We have applied real and synthetic data sets  and presented experimental results that show that our methods are scalable and can recover aggregation functions  that yield high-quality clusterings.  

This paper only scratches the surface of the clustering problem with multiple edge types. There are many  interesting problems to be investigated such as meta-clustering, (i.e., clustering the clusterings) and finding significantly different clusterings for the same data,  which are part of our ongoing work.  We are also planning to extend our experimental work on  the current problem of   computing aggregation functions from ground-truth data. 

\footnotesize
\bibliographystyle{plain}
\bibliography{ali,library}

\end{document}